\documentclass[aps,prl,superscriptaddress,preprint]{revtex4}

\usepackage{bm}
\usepackage{graphicx}
\newcommand\D{{\mathcal D}}
\newcommand\Z{{\mathcal Z}}
\newcommand\bq{{\bm{q}}}
\newcommand\bK{{\bm{K}}}
\newcommand\bA{{\bm{A}}}

\begin{document}

\title{Simulating Quantum Dissipation in Many-Body Systems}

\author{Luca Capriotti}
\altaffiliation[Present address:]{ Institute for Theoretical Physics,
University of California, Santa Barbara, California 93106-4030.}
\affiliation{Istituto Nazionale per la Fisica della Materia (INFM),
  Unit\`a di Ricerca di Firenze, Via G. Sansone 1, I-50019 Sesto
  Fiorentino (FI), Italy}

\author{Alessandro Cuccoli}
\affiliation{Dipartimento di Fisica dell'Universit\`a di Firenze,
    Via G. Sansone 1, I-50019 Sesto Fiorentino (FI), Italy}
\affiliation{Istituto Nazionale per la Fisica della Materia (INFM),
  Unit\`a di Ricerca di Firenze, Via G. Sansone 1, I-50019 Sesto
  Fiorentino (FI), Italy}

\author{Andrea Fubini}
\affiliation{Dipartimento di Fisica dell'Universit\`a di Firenze,
    Via G. Sansone 1, I-50019 Sesto Fiorentino (FI), Italy}
\affiliation{Istituto Nazionale per la Fisica della Materia (INFM),
  Unit\`a di Ricerca di Firenze, Via G. Sansone 1, I-50019 Sesto
  Fiorentino (FI), Italy}

\author{Valerio Tognetti}
\affiliation{Dipartimento di Fisica dell'Universit\`a di Firenze,
    Via G. Sansone 1, I-50019 Sesto Fiorentino (FI), Italy}
\affiliation{Istituto Nazionale per la Fisica della Materia (INFM),
  Unit\`a di Ricerca di Firenze, Via G. Sansone 1, I-50019 Sesto
  Fiorentino (FI), Italy}

\author{Ruggero Vaia}
\affiliation{Istituto di Elettronica Quantistica del Consiglio
Nazionale
    delle Ricerche, Via Panciatichi~56/30, I-50127 Firenze, Italy}
\affiliation{Istituto Nazionale per la Fisica della Materia (INFM),
  Unit\`a di Ricerca di Firenze, Via G. Sansone 1, I-50019 Sesto
  Fiorentino (FI), Italy}

\date{\today}

\begin{abstract}
  An efficient Path Integral Monte Carlo procedure is proposed to
  simulate the behavior of quantum many-body dissipative systems
  described within the framework of the influence functional.
  Thermodynamic observables are obtained by Monte Carlo sampling of
  the partition function after discretization and Fourier
  transformation in imaginary time of the dynamical variables.  The
  method is tested extensively for model systems, using realistic
  dissipative kernels.  Results are also compared with the predictions
  of a recently proposed semiclassical approximation, thus testing the
  reliability of the latter approach for weak quantum coupling. Our
  numerical method opens the possibility to quantitatively describe
  real quantum dissipative systems as, e.g., Josephson junction
  arrays.
\end{abstract}



\maketitle


The description of open quantum systems has significant implications in
the fundamental concepts of quantum mechanics.  It involves issues as
the breakdown of the Schr\"odinger picture and the connection between
the `reversible' and the `irreversible' world~\cite{zurek}. These basic
problems gave the first strong boost to the research in this field in
the early 50's and still are among the most fascinating aspects of the
topic.

In the recent years a renewed interest in the study of quantum
dissipation has come mainly from condensed-matter physics, as systems
with an intermediate ({\em mesoscopic}) scale have been experimentally
developed and theoretically analyzed.  The fundamental issues in
quantum and statistical mechanics are closely tied to technical
applications, such as the single electron transistors~\cite{natureSET},
the Josephson quantum-bits~\cite{natureQbit}, the
SQUIDS~\cite{natureSquid}, and many others. In such systems, the
characteristic quantum effects involve a macroscopic number of
particles. The sizeable dimension of the devices implies that the
relevant dynamical variables can couple to a very large (infinite)
number of degrees of freedom of the surrounding {\em environment} (or
{\em dissipation bath}).  The interaction between the open system and
its environment leads in general to dissipation, fluctuation,
decoherence and irreversible processes. These result in dramatic
changes in the behavior of the system, for instance the dissipative
phase transition in Josephson junction arrays~\cite{yama}.

At variance with classical thermodynamics, which is unaffected by
dissipation, quantum thermodynamics is an ideal field to study the
genuine interplay between quantum fluctuations and dissipation, which
leads in general to interesting physics in the regimes of high quantum
coupling and/or low temperature.  Unfortunately, a suitable theoretical
approach, allowing a faithful comparison with the experimental findings
in the regime of high quantum coupling, is still lacking.  In this
Letter, we present an original Path Integral Monte Carlo (PIMC)
approach, which allows one to efficiently deal with the quantum
thermodynamics of many-body open systems in a wide range of couplings
and temperature; we will show how working with Fourier transformed
variables, and taking advantage of our knowledge of the exact quantum
propagator of harmonic systems, allows us to get reliable results for
many body-systems with reasonable numerical efforts.

We address the problem of quantum dissipation within the
Caldeira-Leggett(CL) framework~\cite{CaldeiraL}, where dissipation is
described as the result of a linear coupling of the physical system of
interest with a bath of harmonic oscillators. By generalizing the CL
formalism to many-body systems, the partition function of a quantum
dissipative system is given by the path integral
\begin{equation}
 \Z=\oint\D[\bq]~e^{-S[\bq]}~,
\label{e.Z}
\end{equation}
with the Euclidean action
\begin{equation}
 S[\bq] =\int_0^{\beta\hbar}\frac{du}\hbar
 \left[ \frac12\, \dot \bq(u)\,\bA~\dot\bq(u)
 + V\big(\bq(u)\big) \right] + S_{\rm{nl}}[\bq]~.
\label{e.S}
\end{equation}
Here, $\bq\equiv\{q_i\}_{i=1,...,M}$ denotes the vector whose
components are the $M$ coordinates of the investigated system and
$\bA\equiv\{A_{ij}\}$ is the mass matrix; the effects of dissipation
are contained in the influence action,
\begin{equation}
 S_{\rm{nl}}[\bq] = \frac{1}{2\hbar}\int_0^{\beta\hbar}\!\!\! du
 \int_0^{\beta\hbar}\!\!\! du'\,\bq(u)\,\bK(u{-}u')~\bq(u')~.
\end{equation}
Within the many-body CL formalism the kernel
$\bK(u)\equiv\{K_{ij}(u)\}$ is an $M{\times}M$ matrix, which is in
general nonlocal in space, i.e., nondiagonal. Thus the dissipation bath
can drive also the spatial correlations of the system as it is
expected, for instance, in the case of shunted Josephson junction
arrays. The kernel matrix $\bK(u)$, depends on the temperature
$T=(k_{{}_{\rm{B}}}\beta)^{-1}$, is a symmetric and periodic function
of the imaginary time $u$, $\bK(u)=\bK(-u)=\bK(\beta\hbar-u)$, and has
a vanishing average over a period.

In order to numerically evaluate the integral appearing in
Eq.~(\ref{e.Z}), the standard PIMC method divides the imaginary-time
interval $[0,\beta\hbar]$ into $P$ slices of width
$\varepsilon=\beta\hbar/P$, $P$ being the so called {\em Trotter
number}. The coordinates $\bq(u)$ turn into the discrete quantities
$\bq_\ell=\bq(\ell\varepsilon)$, $\dot\bq(u)\to
P(\beta\hbar)^{-1}(\bq_\ell-\bq_{\ell-1})$, and the partition function
$\Z$ is obtained as the $P\to\infty$ extrapolation of
\begin{equation}
 {\cal Z}_P= C \beta^{-\frac{MP}{2}}\prod_{i=1}^{M} \int dq_{i0}
 \int \prod_{\ell=1}^{P-1}dq_{i\ell}~ e^{-S_P[\{\bq_\ell\}]}~,
\label{Zp.time}
\end{equation}
where $S_P[\{\bq_\ell\}]$ is the discretized form of the
action~(\ref{e.S}) and $C$ is a temperature-independent normalization;
the related macroscopic thermodynamic quantities are obtained through
the accordingly generated estimators.

The application of the standard PIMC approach to dissipative systems is
made difficult by the fact that the kernel $\bK(u{-}u')$ is usually not
explicitly known in the imaginary-time domain, but rather in the
Matsubara frequency one. In fact, it is generally given in terms of the
bath spectral density or in terms of the Laplace transform of the
damping function appearing in the phenomenological Langevin
equation~\cite{Weiss99}. This makes $\bK(u{-}u')$ long-ranged,
cumbersome to evaluate and sometimes ill-defined as, e.g., in the case
of the widely used Ohmic (or Markovian) dissipation. A possible way to
avoid such problem could be the use of Fourier path-integral
approaches, possibly supported by the partial-averaging
scheme~\cite{FPIJChem}: the latter method, however, still deals with
the problem of evaluating path averages of the potential over
continuous paths.

What we propose here is to start from the finite-$P$ expression
(\ref{Zp.time}) of the standard PIMC for the partition function and
make there a lattice (discrete) Fourier transform, changing the
integration variables from $q_{i\ell}$ to $q_{ik}$ by setting:
\begin{equation}
 \bq_\ell=\bar\bq+\sum_{k=1}^{P-1}\bq_{k}~ e^{i\,2\pi\ell k/P}
\end{equation}
so that:
\begin{eqnarray}
 {\cal Z}_P&=& C\beta^{-{PM\over 2}}\prod_{i=1}^{M}\int d\bar
 q_i\int \prod_{\ell=1}^{P-1}dq_{ik}
\cr
 && \exp\Bigg\{ -\sum_{k=1}^{P-1} \bq_k\bigg[
 \frac{2P^2}{\beta\hbar^2}~\sin^2\frac{\pi k}P \bA +
 \frac{\beta}{2}\bK_k \bigg]\bq_k^* -
\cr
 &&\hspace{35mm}
 -\frac\beta P~\sum_{\ell} V\big(\bq_\ell \big)\Bigg\}~,
\label{Zp}
\end{eqnarray}
where $\bK_k\equiv\{K_{ij,k}\}$ is the usual Matsubara transform
$\bK_k=\int_0^{\beta\hbar}du~\bK(u)~e^{-i\nu_k
  u}$ of the dissipative kernel matrix at the Matsubara frequency $\nu_k=2\pi
k/\beta\hbar$.

We must however observe that the change of variables above is not
enough to get a really efficient procedure to simulate many-body
systems. In fact, in order to get a reliable thermodynamic limit
finite-size effects have to be negligible; as a consequence the number
$M$ must be large enough, so that reaching high values of $P$ may
become computationally very demanding, making the extrapolation to
$P\to\infty$ problematic. However, such difficulty can be largely
circumvented by making use of our knowledge of both the finite- and
infinite-$P$ exact partition function of pure bilinear
actions~\cite{CMPTV95}.  According to Ref.~\cite{CMPTV95} [Eqs.~(38)
and (44)] any Monte Carlo estimate $G(P)$ of a given quantity $G$
obtained at finite $P$ can be corrected by adding the exact
($P\to\infty$) value $G^{(h)}_{\rm HA}$ and subtracting the finite-$P$
estimate $G^{(h)}_{\rm HA}(P)$ of the same quantity~\cite{note} for
the (self-consistent) harmonic approximation of the dissipative
action, getting:
\begin{equation}
G_{\rm HA}(P)=G(P)+\left[G^{(h)}_{\rm HA}-G^{(h)}_{\rm HA}(P)\right]~.
\label{harmsub}
\end{equation}
As it will clearly appear in the following applications, the last step
turns out to be essential and truly effective in the investigation of
many-body systems.

As a preliminary test, we take a single particle
of mass $m$ in a non-linear potential $V(q)=\epsilon{v}(q/\sigma)$,
where $\epsilon$ and $\sigma$ define the energy and length scale,
respectively. In order to better examine the combined effects of
quantum fluctuations and dissipation, it is useful to introduce the
reduced temperature $t=(\beta\epsilon)^{-1}$ and the quantum coupling
$g=\hbar\omega_0/\epsilon$, where
$\omega_0=\sqrt{\epsilon{c^2}/m\sigma^2}$ and $c^2=v''(x_{\rm{m}})$,
$x_{\rm m}$ being the absolute minimum of $v(x)$. For odd Trotter
number, $P=2N+1$, we get:
\begin{eqnarray}
 {\cal Z}_P = C t^{P\over 2}\int d\bar x\int \prod_{k=1}^N da_kdb_k
 ~\exp\Bigg\{\!-\frac{1}{t P} ~ \sum_{\ell}v(x_\ell)+ &&
\cr
 - \sum_{k=1}^N \bigg[
 \frac{4c^2tP^2}{g^2}~\sin^2\frac{\pi k}P
 + \frac{c^2}{t} \kappa_k \bigg](a_k^2+b_k^2)\Bigg\}\,,~~&&
\label{Zp.realodd}
\end{eqnarray}
where $x_\ell=\bar{x}+2\sum_{k=1}^N\big(a_k \cos\frac{2\pi{k}\ell}P +
b_k\sin\frac{2\pi{k}\ell}P\big)$, $\kappa_k=K_k/\omega_0^2$ and we have
used the symmetry properties of $K(u)$, so that
$\kappa_{P-k}=\kappa_k$. The real Fourier variables $\bar{x}$, $a_k$
and $b_k$ are dimensionless, and the integrals in
Eq.~(\ref{Zp.realodd}) may be numerically evaluated by standard Monte
Carlo sampling techniques, e.g. the Metropolis one.

\begin{figure}
\includegraphics[bbllx=14mm,bblly=94mm,bburx=173mm,bbury=207mm,
    width=80mm,angle=0]{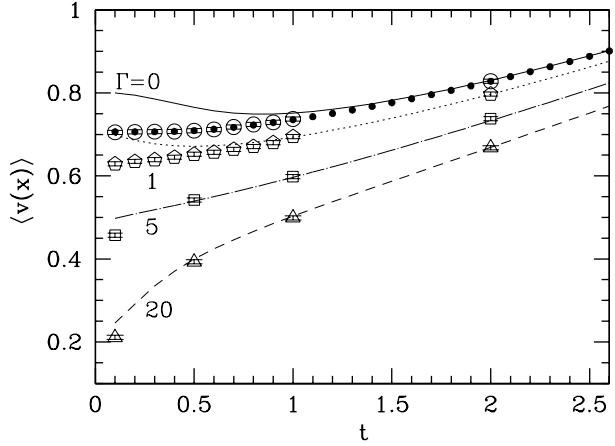}
\caption{\label{vq}
 Temperature dependence of the average potential energy
 $\langle{v({x})}\rangle$ for the single particle in a quartic double
 well, for $g=5$ and different values of the damping strength $\Gamma$.
 Empty symbols are PIMC data, lines are PQSCHA predictions
 (Ref.~\protect\cite{CRTV97}) and filled circles are the exact results
 for $\Gamma=0$.  }
\end{figure}

Fig.~\ref{vq} shows the results obtained for the average potential
energy when a quartic double-well potential $v(x)=(1-x^2)^2$ and Ohmic
dissipation, i.e., $\kappa_k=2\pi(t/g)\Gamma{k}$, are considered
($\Gamma$ is the damping strength in units of $\omega_0$); the same
model was already investigated in Ref.~\cite{CRTV97} by means of the
{\it pure-quantum self-consistent harmonic approximation} (PQSCHA), a
semiclassical approach.  Data in Fig.~\ref{vq} refer to $g=5$ and
different values of damping.  The reported Monte Carlo data represent
the extrapolation to $P\to\infty$ of the results obtained at
$P=17,33,65$, and $129$; the reliability of the algorithm is proven by
the perfect agreement between the exact results and the PIMC data in
the non-dissipative system ($\Gamma=0$). For the dissipative model, the
PIMC data provide therefore a reference, still lacking until now, to
verify the validity of the PQSCHA~\cite{CFTV99}.

Let us turn now to a true many-body dissipative system, and consider
the quantum discrete $\phi^4$ chain, whose Hamiltonian may be written
as~\cite{CFTV99}
\begin{eqnarray}
 \hat{\cal H}&=& \varepsilon_{{}_{\rm{K}}} \bigg[\,
 \frac{Q^2R}3\sum_{i=1}^M \hat p_i^2+V(\hat\bq)\, \bigg]~,
\label{e.Hphi4}
\\
 V(\bq)&=& \frac{3}{2R}\sum_{i=1}^M\bigg[v(q_i)
 +\frac{\,R^2}2\,(q_i-q_{i-1})^2\bigg] ~,
\label{e.Vphi4}
\end{eqnarray}
where $v(x)=(1-x^2)^2/8$, $Q$ is the quantum coupling, and
$\varepsilon_{{}_{\rm{K}}}$ and $R$ are the kink energy and length,
respectively, in the classical continuum limit.

Assuming identical independent baths~\cite{CFTV99} for each degree of
freedom, i.e. $K_{ij}(u)=\delta_{ij}K(u)$, Eq.~(\ref{Zp}) gives
\begin{equation}
 {\cal Z}_P = C t^{PM\over 2}\displaystyle{\prod_{i=1}^M
 \int d\bar q_i \int \prod_{k=1}^N da_{ik}\,db_{ik}~ e^{-S_P}~},
\label{Zp.phi4}
\end{equation}
where we set $q_{i\ell}=\bar{q}_i+2\sum_{k=1}^N\big(a_{ik}
\cos\frac{2\pi{k}\ell}P+b_{ik}\sin\frac{2\pi{k}\ell}P\big)$, again with
$P=2N+1$; using the dimensionless temperature
$t=(\beta\varepsilon_{{}_{\rm{K}}})^{-1}$, the discretized action reads
\begin{eqnarray*}
 S_P=\sum_{i=1}^M\Bigg\{\sum_{k=1}^N \bigg[ \frac{6 t P^2}{Q^2
 R}~\sin^2\frac{\pi k}P + \frac{3}{2 R t} K_k \bigg]
 (a_{ik}^2+b_{ik}^2){+}&&
\cr
 {+}\frac {3R}{4t} \Big[ \! (\bar q_i{-}\bar q_{i-1})^2
 {+}2\sum_{k=1}^N \! \big[
 (a_{ik}{-}a_{i-1k})^2\!{+}(b_{ik}{-}b_{i-1k})^2\big]\! \Big]{+}&&
\cr
 +\frac{3}{2RtP}~\sum_{\ell=1}^P v\left(q_{i\ell}\right)\Bigg\}~.~~&&
\end{eqnarray*}

\begin{figure}
\includegraphics[bbllx=14mm,bblly=94mm,bburx=173mm,bbury=207mm,
    width=80mm,angle=0]{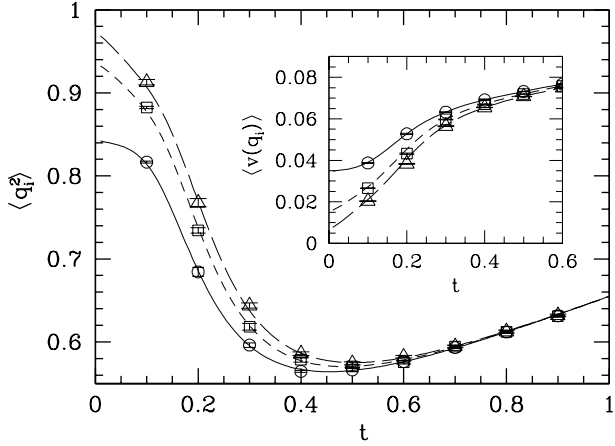}
\caption{\label{q2vq}
 $\langle{q}_i^2\rangle$ and $\langle{v}({q}_i)\rangle$ (inset) vs
 temperature for the $\phi^4$ chain, with $Q=0.2$, $R=5$,
 $\Omega_{{}_{\rm{D}}}=100$ and different values of $\Gamma$.  Empty
 symbols are PIMC data ($P\to\infty$ extrapolations) and lines are
 PQSCHA predictions (Ref.~\protect\cite{CFTV99}). $\Gamma=0$: circles
 and solid line; $\Gamma=20$: squares and short-dashed line;
 $\Gamma=100$: triangles and long-dashed line.  }
\end{figure}

\begin{figure}
\includegraphics[bbllx=14mm,bblly=94mm,bburx=178mm,bbury=207mm,
    width=80mm,angle=0]{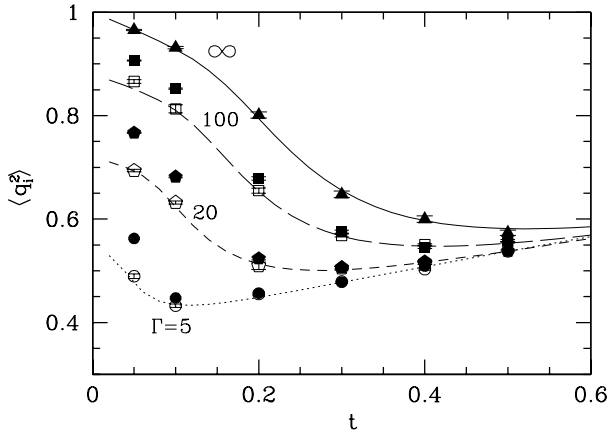}
\caption{\label{q2}
 $\langle{q}_i^2\rangle$ vs temperature for the $\phi^4$ chain, with
 $Q=1$, $R=3$, $\Omega_{{}_{\rm{D}}}=10$ and different values of
 $\Gamma$. Full symbols are PIMC data at finite Trotter number ($P=81$,
 for finite $\Gamma$, $P=11$ for $\Gamma\to\infty$); empty symbols are
 $P\to\infty$ extrapolations. Lines are PQSCHA predictions.  }
\end{figure}

\begin{figure}
\includegraphics[bbllx=17mm,bblly=64mm,bburx=174mm,bbury=237mm,
    width=80mm,angle=0]{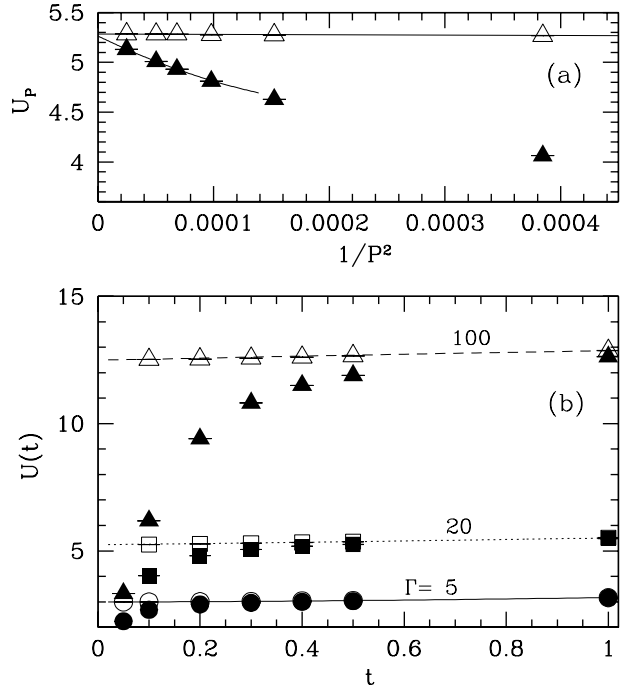}
\caption{\label{erg}
 Internal energy (per site) $U$ for the $\phi^4$ chain, with $Q=1$,
 $R=3$, $\Omega_{{}_{\rm{D}}}=10$, and different values of $\Gamma$.
 (a): $P$-dependence of $U$ for $\Gamma=20$, and $t=0.2$. Full symbols:
 bare PIMC results $U(P)$; empty symbols: harmonically-corrected data
 $U_{\rm{HA}}(P)$. Lines are quadratic fits. (b): temperature dependence
 of $U$. Full symbols: PIMC data for $P=101$; empty symbols:
 harmonically-corrected $P\to \infty$ extrapolations. Lines are PQSCHA
 predictions.  }
\label{uenphi4}
\end{figure}

The average quantities for the dissipative $\phi^4$ chain presented in
Figs.~\ref{q2vq}--\ref{uenphi4} have been obtained for periodic chains
of length ($\sim 10^2$ sites) large enough to be representative of the
thermodynamic limit for each set of physical parameters and by
extrapolating to $P\to \infty$ the results given by simulations at
finite $P$.  A Drude-like spectral density was assumed for the
environmental interaction~\cite{Weiss99,CFTV99}, so that the
dissipative kernel reads
\begin{equation}
 K_k=\Gamma\Omega_{{}_{\rm{D}}}
 ~\frac{(2\pi t k/Q)}{\Omega_{{}_{\rm{D}}}+(2\pi t k/Q)}~,
\label{drudekern}
\end{equation}
where the dissipation strength $\Gamma$ and the cut-off frequency
$\Omega_{{}_{\rm{D}}}$ are measured in units of
$\Omega=Q\varepsilon_{{}_{\rm{K}}}$.  Comparison of the PIMC results
with those of the PQSCHA~\cite{CFTV99}, shown in Fig.~\ref{q2vq},
clearly indicates that the predictions of the latter are impressively
accurate. Remarkably, the accuracy is also preserved for fairly large
values of quanticity, quite close to the predicted limits of
applicability of the PQSCHA scheme, as it appears in Fig.~\ref{q2}.

The role played by the correction scheme of Eq.~(\ref{harmsub}) is
clearly shown in Fig.~\ref{uenphi4}~(a), where $U_{\rm HA}(P)$ displays
a very weak dependence on $P$, thus allowing to get reliable estimates
of the $P\to\infty$ extrapolation even by starting from results
obtained for relatively small values of $P$.  The relevance of the
harmonic correction is more and more evident for increasing dissipation
strength [Fig.~\ref{uenphi4}~(b)]: indeed, in this regime the
finite-$P$ bias essentially comes from the bilinear contribution of the
influence functional and the latter can be easily healed by the above
correction scheme, so that, for instance, the characteristic almost
flat behavior of $U$ at low temperature~\cite{CFTV99} is obtained
without using exceedingly high values of $P$.

In conclusion, we have introduced a formulation of the Path Integral
Monte Carlo method which greatly simplifies the numerical investigation
of the thermodynamics of quantum dissipative systems with bilinear
influence action, making it possible to afford also many-body systems.
The combined use of both the discrete Fourier transformed
representation of the dynamical variables and the correction scheme of
Eq.~(\ref{harmsub}) is essential to efficiently circumvent the
numerical difficulties arising from the non locality in time and space
of the action describing an open many-body system. The application to a
double-well potential and to a $\phi^4$ chain have shown the power of
the method and have given, as a byproduct, a direct confirmation of the
validity of the semi-classical approach of Refs.~\cite{CRTV97,CFTV99}
in the expected parameter region, i.e., weak quantum coupling and/or
strong dissipation. The numerical approach introduced here does not
suffer of such limitations, and we expect it to be useful for getting
meaningful insight on the behavior of strongly quantum systems in
presence of dissipation, thus making it possible to address interesting
problems of mesoscopic physics as, e.g., the dissipative transition in
Josephson junction arrays~\cite{yama}.

\begin{acknowledgments}
  Valuable discussions with W. Janke, T. Sauer, U.
  Weiss and J. Stockburger are gratefully acknowledged.  This work was
  supported by MURST (COFIN00) and CRUI (Vigoni program); A.F. thanks
  ``Fondazione Della Riccia'' for financial support and the Institut
  f\"ur Theoretische Physik II, Universit\"at Stuttgart, for kind
  hospitality.  
\end{acknowledgments}



\begin{thebibliography}{99}
\bibitem{note} For a quadratic action with CL dissipation, any
  thermodynamic quantity at finite $P$ can be exactly obtained
  starting from the density matrix given by Eqs.~(A14),~(36), and~(37)
  of Ref.~\protect{\cite{CFTV99}}, with $w=0$ and $\infty$ replaced by
  $N$ (as defined in Eqs. (\ref{Zp.realodd}) and (\ref{Zp.phi4}) ) in
  the summation limits.
\end{thebibliography}
\end{document}